# Transmutation of Minor Actinides and Power Flattening in PWR MOX Fuel

Shengli Chen[a] and Cenxi Yuan[a,*]

[a]*Sino-French Institute of Nuclear Engineering and Technology, Sun Yat-sen University, Zhuhai, Guangdong 519082, China*

[*]Corresponding author: yuancx@mail.sysu.edu.cn

## 1. Introduction

The management of long-lived radioactive products in the spent nuclear fuel is one of the most difficult issues in nuclear engineering. The closed cycle of spent fuel and the transmutation of some long-lived radioactive nuclides have been proposed as an alternative concept for solving the problem.

The uranium-plutonium Mixed OXide (MOX) fuel was studied in the past decades in some European countries and Japan [1] in order to recycle the plutonium, which is about 1% in the spent fuel [2]. Some plutonium isotopes have long half-life and large fission cross section. Accordingly, they can be used to produce energy by fission, which can also reduce long-lived radioactive isotopes. MOX fuel is expected to be used more and more in order to reduce the spent fuel storage burden. This is the trend of the nuclear power generation in the Chinese utility companies; for example, the MOX fuel may be used in Taishan Evolutionary Power Reactor (EPR) in the future.

The Minor Actinides (MAs) have very long decay half-lives among all radioactive nuclides in the spent fuel. Accordingly, the study of MAs transmutation is a significant work for the post-processing of spent fuel. The transmutation of MAs is studied in Pressurized Water Reactor (PWR) [3][4][5], Fast Reactor (FR) [5][7][8] and subcritical reactors [9]. Nowadays, the majority of commercial nuclear power reactors operating in the world are PWRs. It is thus important to investigate the transmutation of MAs in the operating PWRs, which provides a potential approach to reduce the inventory of high level long-lived radioactive MAs in the world.

The $^{237}$Np is one of the most principal MAs in the depleted nuclear fuel of PWRs. The percentage of this nuclide in the total MAs can arrive 56.2% [4]. In addition, it has the longest decay half-life among MAs. Hence, it is important to study the transmutation of $^{237}$Np loading in addition to the transmutation of a composition of MAs.

In the previous studies in PWR, Chen *et al.* [3] have analyzed the transmutation in PWR MOX fuel with homogeneous $^{237}$Np and mixed MAs loading. Liu *et al.* [4] focused on both homogeneous and heterogeneous MAs distribution in a reactor core; Hu *et al.* [5] studied the heterogeneous core with MAs coated poison rods.

The heterogeneous MAs loading is excellent in the consideration of spatial self-shielding and the reactivity control. In addition, the heterogeneous power distribution in a PWR fuel assembly exists due to the distribution of the moderator. The study in Ref.[3] has proved the decrement of reactivity by adding MAs in PWR MOX fuel. In this context, we suggest a semi-heterogeneous loading of MAs in PWR MOX fuel so that we can transmute MAs and flatten power distribution in an assembly. The semi-heterogeneous MAs loading is that we load homogeneously MAs in fuel rods, but MAs loaded fuel rods are heterogeneously distributed in an assembly. The reason is that the homogeneous loading is easier for the fuel fabrication and the heterogeneous distribution of loaded fuel rods can flatten the power distribution. The MOX fuel is considered because both the plutonium in MOX fuel and the MAs are extracted in the depleted nuclear fuel. It is thus not necessary to completely separate them in the post-processing of the depleted fuel.

The present work studies the transmutation and the effect on the power flattening. We focus on 3%wt $^{237}$Np loading and 3%wt MAs composition loading in 92 MAs loaded rods in both high and low concentration MOX fuels. Finally, the effective multiplication factor $k$ is compared among different cases.

## 2. Methodology

The typical 17×17 PWR assembly design is used in the present study, as shown in Fig. 1. The 24 guide tubes and the central instrumentation tube are full of the moderator when there is no insertion of control rods or instrument, which is the case for present work. In the cases of MAs loading, 92 fuel rods around tubes (orange rods in Fig. 1) are replaced by MAs loaded fuel rods. All simulations are carried out on an assembly. The geometrical parameters are the same as listed in Ref.[3] and Ref.[11]. The simulations in the present work are performed using the Monte Carlo code RMC [12], which is a 3-D Monte Carlo neutron transport code developed by Tsinghua University. As a Monte Carlo code, the RMC is able to deal with complex geometry. In addition, this code can use continuous energy pointwise ENDF/B-VII.0-based cross sections for different materials and at different temperatures. It has both criticality and burnup calculations, which can obtain the effective multiplication factor and the nuclide concentrations at different burnup level.

Two sets of MOX fuel are considered in the present work to compare the transmutation and power flattening efficiency with different plutonium concentrations in MOX fuel. The high (low resp.) concentration MOX fuel contains 9.8% (6.6% resp.) plutonium isotopes in the fuel. The enrichment of each isotope in MOX fuel is shown in Table I.

In general, the decay half-life of MA is very long. Among MAs nuclides, the most enrich nuclide in the spent fuel $^{237}$Np has the longest decay half-life $2.144\times10^6$ years. After the $^{237}$Np, the $^{245}$Cm and $^{243}$Am have the longest decay half-life, which is 8500 years and 7070 years respectively. Another enriched nuclide in the spent fuel is $^{241}$Am, which half-life is 432 years. However, this is not the case for $^{242}$Am due to its high thermal neutron fission cross section and its short half-life (16 hours). It is thus not necessary to transmute $^{242}$Am in the MOX fuel.

Due to the long half-live of $^{237}$Np and the large quantity in spent fuels, its transmutation is of large importance for the post-processing of spent fuel. In this context, loading of 3% $^{237}$Np in 92 fuel rods is studied in the present work. The initial quantity of MAs in 3% MAs loaded 92 fuel rods is almost the same as that in 1% MAs homogeneously loaded assembly (i.e. 264 rods).

In addition, in order to transmute other long-live MAs, it is also of great interest to study the mixed long-lived MAs. In the case of 3% mixed MAs loading in two concentrations of MOX fuels, the percentage of each MA nuclide is shown in Table II. This corresponds to the relative percentage of these 5 MAs in the 10 years cooled spent fuel unloaded from a 3 GW thermal power reactor at 30 GWd/t [13].

much higher power due to the larger moderator to fuel ratio. The ratio of maximum power to minimum power is 1.28. Accordingly, we replace these fuel rods with high power by MAs loaded fuel rods, as shown in Fig. 1, to transmute MAs and to flatten power distribution.

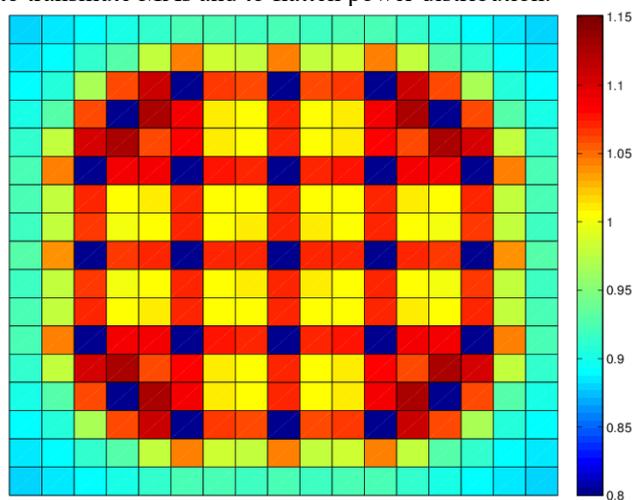

Fig. 2. Relative power distribution in high concentration MOX assembly at BOF

*3.1 Transmutation of MAs*

In the present work, to study the transmutation for each isotope, we define 3 quantities: net transmutation rate, noted as $R_N$; total transmutation rate $R_T$; and the equivalent natural decay time Te. $R_N$ defines the percentage of transmuted concentration to initial loading concentration. $R_T$ is the transmutation rate by considering the production of MA in an unloaded assembly. Te represents the equivalent decay time to achieve the $R_T$ by natural decay.

- $R_N = (C_0 - C_2)/C_0 \times 100\%$  (1)
- $R_T = (C_0 - C_2 + C_1)/C_0 \times 100\%$  (2)

where $C_0$ represents the loading concentration of the MA, $C_1$ and $C_2$ are concentrations of the MA after depletion in the unloaded case and MAs loaded cases, respectively.

The results of the transmutation for the 3% $^{237}$Np loaded case are shown in Table III. The transmutation of 3 principal MAs in 3% mixed MAs loading case gives the results in Table IV.

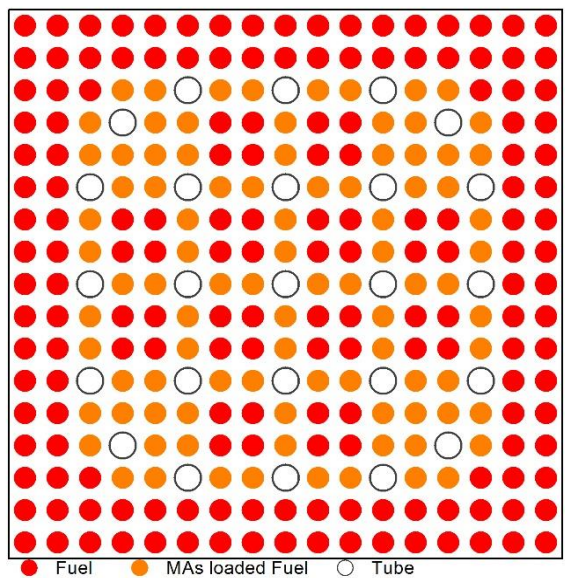

Fig. 1. 17×17 PWR lattice configuration

Table I. Actinide initial enrichment in MOX fuels

|  | $^{235}$U | $^{238}$U | $^{238}$Pu | $^{239}$Pu | $^{240}$Pu | $^{241}$Pu | $^{242}$Pu |
|---|---|---|---|---|---|---|---|
| Low | 0.23 | 93.2 | 0.10 | 3.97 | 1.62 | 0.58 | 0.33 |
| High | 0.23 | 90.0 | 0.15 | 5.90 | 2.40 | 0.86 | 0.49 |

Table II. Percentage of loaded MAs in MOX

| Isotope | $^{237}$Np | $^{241}$Am | $^{243}$Am | $^{244}$Cm | $^{245}$Cm |
|---|---|---|---|---|---|
| wt% | 41.80 | 47.86 | 8.62 | 1.63 | 0.09 |

## 3. Results and discussion

As shown in Fig. 2, the fuel rods around tubes have

Table III. Transmutation of loaded $^{237}$Np in MOX

|  | $R_N$ (%) | $R_T$ (%) | Te (year) |
|---|---|---|---|
| Low MOX | 54.2 | 55.1 | $2.48\times10^6$ |
| High MOX | 46.9 | 47.9 | $2.01\times10^6$ |

Table IV. Transmutation of loaded MAs in MOX

| Fuel | Isotope | $R_N$ (%) | $R_T$ (%) | Te (yr) |
|---|---|---|---|---|
| Low | $^{237}$Np | 52.6 | 54.7 | $2.45\times10^6$ |
|  | $^{241}$Am | 65.7 | 70.1 | 753 |
|  | $^{243}$Am | -- | 24.8 | $2.90\times10^3$ |
| High | $^{237}$Np | 45.4 | 47.7 | $2.00\times10^6$ |
|  | $^{241}$Am | 44.7 | 56.0 | 512 |
|  | $^{243}$Am | -- | 21.0 | $2.41\times10^3$ |

As explained in Ref.[3], the transmutation is more efficient in lower plutonium concentration fuel due to the competition of reaction with neutron between plutonium and MAs. The transmutation efficiency of the $^{241}$Am is the highest due to its large thermal neutron capture cross section, which is 412 barns [14]. However, it is more interesting to study the $^{237}$Np due to its very long half-live.

In the consideration of equivalent natural decay time, the transmutation of $^{237}$Np is the most important. In both $^{237}$Np loading case and mixed MAs loading case, the transmutation efficiency is similar, corresponding to $2.5\times10^6$ ($2.0\times10^6$ resp.) years natural decay in low (high resp.) concentration MOX fuel. This isotope is produced by double neutron capture reaction of $^{235}$U or α decay of $^{241}$Am, but the concentration of the $^{235}$U is very low and the 432 years half-live decay of the $^{241}$Am has very limited effect during a cycle life. Therefore, the addition of other MAs has very limited influence on the transmutation efficiency of $^{237}$Np.

From the consideration of equivalent natural decay time by transmutation, the $^{237}$Np loading is better than mixed MAs loading case because of the long time to reduce $^{237}$Np by natural decay.

*3.2 Power flattening*

One of the most important objectives of this work is the study of power flattening by transmuting MAs in high power fuel rods. We note that no MAs loading case as the reference case, 3% $^{237}$Np (mixed MAs resp.) loading as the case 1 (case 2 resp.). The present work gives the standard deviation σ of relative power and the ratio of maximum power to minimum power at the Begin of Life (BOF, 0 GWd/t) in Table V and Table VI. The same analyses for the burnup at the End of Life (EOF, 50 GWd/t) are also shown in Table V and Table VI for both low concentration and high concentration MOX fuels.

Table V. Deviation of power distribution at BOF and burnup distribution at EOL in low MOX fuel

|  | Power (BOF) |  | Burnup (EOL) |  |
|---|---|---|---|---|
|  | σ | Max/Min | σ | Max/Min |
| Ref. | 7.2% | 1.28 | 5.5% | 1.21 |
| Case 1 | 4.9% | 1.18 | 4.5% | 1.16 |
| Case 2 | 3.0% | 1.13 | 4.5% | 1.15 |

Table VI. Deviation of power distribution at BOF and burnup distribution at EOL in high MOX fuel

|  | Power (BOF) |  | Burnup (EOL) |  |
|---|---|---|---|---|
|  | σ | Max/Min | σ | Max/Min |
| Ref. | 7.2% | 1.28 | 6.1% | 1.23 |
| Case 1 | 5.3% | 1.20 | 4.9% | 1.18 |
| Case 2 | 3.8% | 1.14 | 4.6% | 1.16 |

According to the results in Table V and Table VI, the power distribution has been flatted by loading MAs in high power fuel rods in an assembly. The mixed MAs loading is more efficient than $^{237}$Np loading due to the larger thermal neutron capture cross section of $^{241}$Am (117 barn for $^{237}$Np [14]), of which the percentage in mixed MAs is about 50%.

The flattening of the burnup at EOF is less efficient than the power distribution at BOF. Firstly, even in no MAs loaded assembly, heterogeneous power distribution becomes less important at higher burnup (σ = 5.0% and Max/Min = 1.19 at EOF for high MOX fuel) due to depletion of fissile isotopes. In addition, the neutron capture reaction of fertile MAs gives the fissionable isotopes due to the odd neutron number of products.

By studying the effect on power and burnup flattening, the mixed MAs loading is better than the only $^{237}$Np loading in both high and low MOX fuels.

The present work replaces high power fuel rods by MAs loaded fuel rods to transmute MAs and to flat power distribution. In PWR, the rim effect is evident in both UO$_2$ and MOX fuel [15][16]. Therefore, neglecting the difficulties industrial fabrication, the heterogeneous MAs loading in a fuel rod is another method to transmute long-lived MAs and to flatten power distribution in each fuel rod.

*3.3 Infinite multiplication factor k*

In previous two sections, we have proved the high transmutation efficiency and satisfactory results on power flattening by adding MAs loaded fuel rods into MOX fuel assembly. However, the addition of MAs has the negative contribution on the reactivity. Accordingly, the study of reactivity is of great importance.

The infinite multiplication factors *k* for the low MOX fuel and high MOX fuel are given in Fig. 3 and Fig. 4. Because the capture cross section of the $^{241}$Am is larger than that of the $^{237}$Np, the flattening of power and burnup distribution of mixed MAs loading is more important. However, this conducts also larger decrement of reactivity at low burnup level. In contrast, at the EOL, the reactivity of the mixed loading case is higher than only $^{237}$Np loading case due to the larger consummation of MAs in the mixed loading case ($^{241}$Am has the largest neutron capture cross section among these MAs).

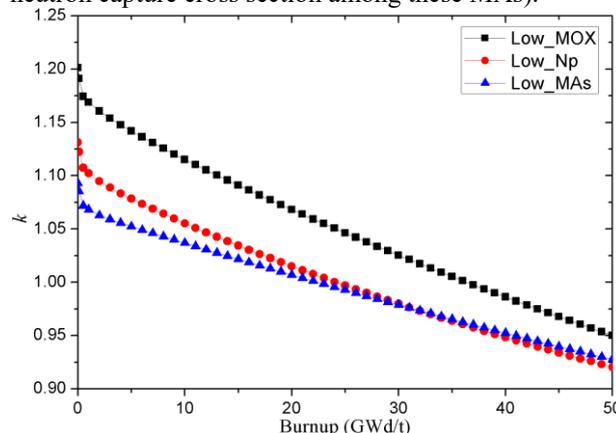

Fig. 3. *k* in low concentration MOX assembly

From the calculation of multiplication factor, we conclude that the mixed MAs loading has less effect on reactivity at the EOL. In addition, compared with $^{237}$Np loading cases, the reactivity varies less for the mixed MAs loading scenarios.

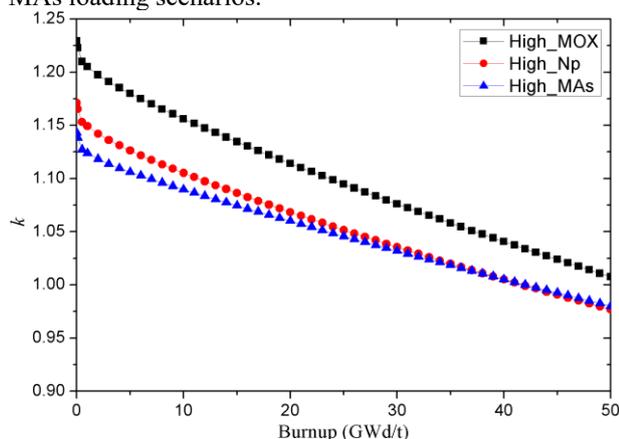

Fig. 4. *k* in high concentration MOX assembly

### 4. Conclusions

In order to transmute long-lived MAs and flatten power distribution in a PWR MOX fuel assembly, the authors have proposed to replace some high power fuel rods by MAs loaded fuel rods. The present work proves the high efficiency of long-lived MAs transmutation in a PWR MOX assembly with 92 fuel rods with 3% MAs loaded. The power and the burnup distribution have been flatted by using the MAs loading method proposed by authors. The $^{237}$Np loading method is expected for the transmutation by comparing with natural decay time to achieve the same reduction. The mixed MAs loading has better behaviors on power flattening and negative contribution of reactivity. In addition, the mixed MAs loading can largely reduce the quantity of $^{241}$Am, while the addition of other MAs has no influence on the transmutation efficiency of $^{237}$Np.

### Acknowledgements

The authors acknowledge the authorized usage of the RMC code from Tsinghua University for this study. This work has been supported by the National Natural Science Foundation of China under Grant No. 11305272, the Special Program for Applied Research on Super Computation of the NSFC Guangdong Joint Fund (the second phase), the Guangdong Natural Science Foundation under Grant No. 2014A030313217, the Pearl River S&T Nova Program of Guangzhou under Grant No. 201506010060, theTip-top Scientific and Technical Innovative Youth Talents of Guangdong special support program under Grant No. 2016TQ03N575, and the Fundamental Research Funds for the Central Universities under Grant No. 17lgzd34.

### References

[1] Frank N. von Hippel, "Plutonium and Reprocessing of Spent Nuclear Fuel," *Science*, **293**, 2397 (2001).

[2] Anastasov V, Betti M, Boisson F, et al., "Status of minor actinide fuel development," *IAEA Nuclear energy series* **J**. NF-T-4.6, 2009.

[3] Chen S, Yuan C, Wu J, et al., "Study of Minor Actinides Transmutation in PWR MOX fuel," *Proc. of the 25$^{th}$ International Conference on Nuclear Engineering*, Shanghai, China, July 2–6, 2017.

[4] Liu B, Wang K, Tu Ji, et al., "Transmutation of minor actinides in the pressurized water reactors," *Annals of Nuclear Energy*, **64**, 86 (2014).

[5] Hu W, Liu B, Ouyang X, et al., "Minor actinide transmutation on PWR burnable poison rods," *Annals of Nuclear Energy*, **7**, 74 (2015).

[6] Nishihara K, Oigawa H, Nakayama S, et al., "Impact of partitioning and transmutation on high-level waste disposal for the fast breeder reactor fuel cycle," *Journal of Nuclear Science and Technology*, **47**, 1101 (2010).

[7] Meiliza Y, Saito M, Sagara H., "Protected plutonium breeding by transmutation of minor actinides in fast breeder reactor," *Journal of Nuclear Science and Technology*, **45**, 230 (2008).

[8] Wakabayashi T, "Transmutation Characteristics of MA and LLFP in a Fast Reactor", *Progress in Nuclear Energy*, **40**, 457 (2002).

[9] Beller D E, Van Tuyle G J, Bennett D, et al., "The U.S. accelerator transmutation of waste program," *Nuclear Instruments and Methods in Physics Research A,* **463**, 468 (2001).

[10] Herrera-Martinez A, Kadi Y, Parks G, "Transmutation of nuclear waste in accelerator-driven systems: Thermal spectrum," *Annals of Nuclear Energy*, **34**, 550 (2007).

[11] Chen S, Yuan C, "Neutronic Analysis on Potential Accident Tolerant Fuel-Cladding Combination U$_3$Si$_2$-FeCrAl," *Science and Technology of Nuclear Installations*, **2017**, 1 (2017).

[12] Wang K, Li Z, She D, et al., "RMC-A Monte Carlo code for reactor physics analysis," *Proc. International Conference on Mathematics and Computational Methods Applied to Nuclear Science and Engineering, M and C*, Sun Valley, ID, USA, May 5-9, 2013, Vol. 1, p. 89-104 (2013).

[13] Broeders C H M, Kiefhaber E, Wiese H W, "Burning transuranium isotopes in thermal and fast reactors," *Nuclear Engineering and Design*, **202**, *2*, 157 (2000).

[14] Iwasaki T, "A Study of Transmutation of Minor Actinide in a Thermal Neutron Field of the Advanced Neutron Source," *Progress in Nuclear Energy,* **40**, 481 (2002).

[15] Yuan C, Wang X, Chen S, "A Simple Formula for Local Burnup and Isotope Distributions Based on Approximately Constant Relative Reaction Rate". *Science and Technology of Nuclear Installation*, **2016**, 6980547 (2016).

[16] Yuan C, Chen S, Wang X, "Monte Carlo Study on Radial Burnup and Isotope Distribution". *Proc. of the 20th Pacific Basin Nuclear Conference*, Beijing, China, April 5-9, 2016, Vol. 3, p. 179-185, Springer, (2016).